\documentclass[12pt,a4,french]{article}

\usepackage{latexsym,amsmath,amssymb,fancybox}

\usepackage{color,epsfig}
\usepackage{graphicx}

\definecolor{rouge}{rgb}{0.8,0.2,0}
\definecolor{violet}{rgb}{0.4,0,0.4}
\definecolor{vert}{rgb}{0,0.6,0.2}
\definecolor{navy}{rgb}{0.0,0.0,0.2}
\definecolor{orange}{rgb}{0.8,0.4,0.0}
\definecolor{brown}{rgb}{0.6,0.2,0.0}
\definecolor{bleu}{rgb}{0.3,0.0,0.8}

\def\kk{{\color{bleu} k}}
\def\hh{{\color{bleu} h}}
\def\ff{{\color{bleu} f}}
\def\KK{{\color{bleu} K}}

\def\rrho{{\color{brown}\rho}}
\def\PP{{\color{brown}P}}
\def\mm{{\color{brown}m}}
\def\aa{{\color{brown}a}}
\def\calD{{\color{black} {\cal D}}}
\def\calJ{{\color{brown} {\cal J}}}
\def\calM{{\color{brown} {\cal M}}}
\def\elle{{\color{brown} \ell}}
\def\calE{{\color{brown} {\cal E}}}
\def\calK{{\color{brown} {\cal K}}}
\def\cc{{\color{brown}c}}
\def\Ome{{\color{brown} \Omega}}

\def\pp{{\color{orange} p }}
\def\Tthet{{\color{orange} \Theta}}

\def\alph{{\color{rouge}\alpha}}
\def\bet{{\color{rouge}\beta}}
\def\pphi{{\color{bleu}\phi}}
\def\phhi{{\color{bleu}\varphi}}
\def\uu{{\color{bleu}u}}
\def\tt{{\color{bleu}t}}

\def\omeg{{\color{violet}\omega}}
\def\XX{{\color{violet}X}}
\def\YY{{\color{violet}Y}}
\def\varth{{\color{violet}\vartheta}}
\def\xx{{\color{violet}x}}
\def\yy{{\color{violet}y}}
\def\zz{{\color{violet}z}}
\def\lle{{\color{violet} n }}
\def\UU{{\color{violet}U}}
\def\varho{{\color{violet}\varrho}}

\def\thet{{\color{rouge}\theta}}
\def\ee{{\color{rouge}e}}
\def\mmu{{\color{rouge}\mu}}
\def\coscr{{\color{rouge}{\cal X}}}

\def\calP{{\color{vert} {\cal P}}}
\def\calR{{\color{vert} {\cal R}}}

\def\rr{{\color{vert} r}}
\def\nn{{\color{vert} n}}
\def\Delt{{\color{vert} \Delta}}
\def\llambda{{\color{vert} \lambda}}

\def\spose#1{\hbox to 0pt{#1\hss}}\def\lta{\mathrel{\spose{\lower 3pt\hbox
{$\mathchar"218$}}\raise 2.0pt\hbox{$\mathchar"13C$}}}  \def\gta{\mathrel
{\spose{\lower 3pt\hbox{$\mathchar"218$}}\raise 2.0pt\hbox{$\mathchar"13E$}}}

\def\Euro{\spose {\lower 2.5pt\hbox{${^{\bf =}}$}}{ C}}

\def\bb{\color{black}  $ }  \def\fb{ $  }
\def\be{\begin{equation} }
\def\fe{\end{equation}}

\def\spose#1{\hbox to 0pt{#1\hss}}\def\lta{\mathrel{\spose{\lower 3pt\hbox
{$\mathchar"218$}}\raise 2.0pt\hbox{$\mathchar"13C$}}}  \def\gta{\mathrel
{\spose{\lower 3pt\hbox{$\mathchar"218$}}\raise 2.0pt\hbox{$\mathchar"13E$}}}

\begin{document}

\title{
{\color{violet}Half century of black-hole theory:\\
from physicists' purgatory to mathematicians' paradise.
\color{orange} }
}

\author { {\bf Brandon Carter,}  \\
{\color{vert}  LuTh, Observatoire Paris-Meudon, France. }
}

\date{\color{navy} Contrib. to 
{\it A Century of Relativity Physics}, Oviedo 2005.}

\maketitle

{\bf Abstract}:  Although implicit in the discovery of the
 Schwarzschild solution 40 years earlier, the issues raised by 
the theory of what are now known as black holes were so unsettling 
to physicists of Einstein's generation that the subject remained 
in a state of semiclandestine gestation  until his demise. That 
turning point  -- just half a century after Einstein's  original 
foundation of relativity  theory, and just half a century ago today  
-- can be considered to mark the birth of black hole theory as a 
subject of systematic development by physicists of a new and
less inhibited generation, whose enthusastic investigations have 
revealed structures of unforeseen mathematical beauty, even though 
questions about the physical significance of the concomitant 
singularities remain controversial.

\vfill\eject

\bigskip\noindent
{\bf 1. Introduction: Schwarzschild's unwelcome solution.}
\medskip

\begin{figure}
\centering
\epsfig{figure=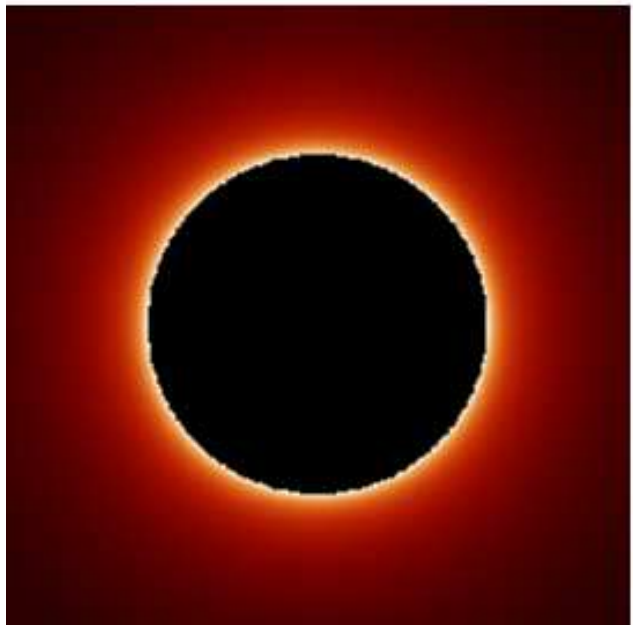, width=8.4cm}
\caption{Numerically simulated view of (isolated) spherical black 
hole illuminated only by uniform distant sky background from 
which light is received only for viewing angle {\bb\alph> \bet\fb} 
where, for an observer falling radially from rest at large distance, 
the angle {\bb\bet\fb} subtended by the hole will be given by  the 
formula (\ref{wholesize}) obtained in the appendix.}
\label{whole}
\end{figure}

This illustrated review is intended to provide a brief overview of the 
emergence, during the last half century, of the theory of ordinary 
(macroscopic 4-dimensional) black holes, considered as a phenomenon 
that (unlike the time reversed phenomenon of white holes) is 
manifestly of astrophysical importance in the real world. The scope 
of this review therefore does not cover quantum aspects such as the 
Bekenstein-Hawking particle creation effect, which is far too weak to 
be significant for the macroscopic black holes that are believed to 
actually exist in the observable universe. Nor does it cover the 
interesting mathematics of higher dimensional generalisations, a subject 
that is (for the time being) so far from relevance to the known physical 
world (in which -- according to the second law of thermodynamics -- the 
distinction between past and future actually matters) that its 
practitioners have formed a subculture in which the senior
members seem to have forgotten (and their juniors seem never to have 
been aware of have been aware of)
the distinction between black and white holes, as they have adopted a 
regretably misleading terminology whereby the adjective ``black'' is 
abusively applied to any brane system that is {\bf hollow} -- including 
the case of an ordinary (black or white) hole, which, to be systematic,
should be classified  as a (black or white) hollow zero brane of 
codimension 3.
 
The rapid general acceptance of the reality and importance of the  
positrons whose existence was implied by  Dirac's 1928 theory
of the electron is in striking contrast with the widespread
resistance to recognition of the reality and importance of the black 
holes whose existence was implied by Einstein's 1915 theory of 
gravity. It is symptomatic that black holes were not even named as
such until more than half a century later. The sloth with which the 
subject has been developed over the years is illustrated by the 
fact that although the simplest black hole solution was already 
discovered (by Schwarzschild) in 1916, the simulation in 
Figure \ref{whole}  of the present review (80 years later) provides
what seems to  the first serious reply to the very easy  
question of what it would actually look like, all by itself, with no 
illumination other than that from a uniform sky background.

Much of the responsibility for the delay in the investigation of
the consequences of his own theory is attributable \cite{Thorne94}
to Einstein himself. Although his work had revolutionary implications,  
Einstein's instincts tended to be rather conservative. It was as a 
matter of necessity (to provide an adequate account first of 
electromagnetism and then of gravitation) rather than preference that 
Einstein introduced  the radically new paradigms involved first in his 
theory of special relativity, just a hundred years ago,  and then in 
the work on general relativity that came to fruition ten years later. 
When cherished prejudices were undermined by the consequences, Einstein
was as much upset as any of his contempory colleagues. It could have 
been said of Albert Einstein (as it was said of his illustrious and like
minded contempory, Arthur Eddington) that he was always profound, but 
sometimes profoundly wrong.

The most flagrant example was occasionned by Friedmann's prescient
1922 discovery of what is now known as the ``big bang'' solution
of the general relativity equations, which Einstein refused to accept
because it conflicted with his unreasonable prejudice in favor of 
a cosmological scenario that would be not only homogeneous (as actually
suggested by subsequently available data) but also static (as commonly 
supposed by earlier generations) despite the incompatibility (in 
thermal disequilibrium) of these alternative simplifications with each 
other and with the obvious observational consideration (known in 
cosmologically minded circles as the Cheseaux-Olbers paradox) that -- 
between the stars -- the night sky is dark.  Einstein's incoherent 
attitude (reminiscent of the murder suspect who claimed to have an
alibi as well as the excuse of having acted in self defense) lead him 
not only to tamper with his own gravitation equations by inclusion of 
the cosmological constant, but anyway to presume without checking that 
Friedmann's (actually quite valid) solution of the original version 
must have been mathematically erroneous.

Compared with his tendency to obstruct progress in cosmology, 
Einstein's conservatism was rather more excusible in the not so simple 
case of what are now known as black holes. It is understandable 
that (like Eddington) he should have been unwilling to explore the 
limitations on the validity of his theory that are indicated by the 
weird and singular -- or as Thorne \cite{Thorne94} puts it 
``outrageous'' -- features that emerge when strong field solutions of 
the general relativity equations are extrapolated too far into the non 
linear regime. 

\begin{figure}
\centering
\epsfig{figure=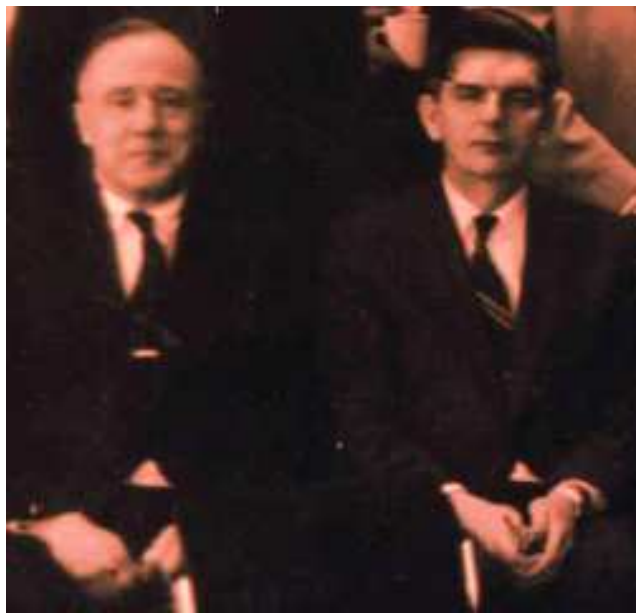, width=8.4 cm}
\caption{John Wheeler with Robert Dicke (Princeton 1971).}
\label{Princeton71}
\end{figure}

At the outset Einstein's interest in the spherical vacuum solution of
his 1915 gravitational field equations was entirely restricted to the 
weak field regime,  far outside the ``horizon'' at {\bb \rr=2\mm\fb} in 
the simple exact solution
{\be {\rm d}s^2= \rr^2({\rm d}\thet^2+{\rm sin}^2\thet\, {\rm d}\pphi^2)
+{\rm d}\rr^2/(1\!-\!{2\mm}/{\rr})
-(1\!-\!{2\mm}/{\rr}){\rm d}\tt^2\!\, ,\fe}
that was obtained within a year, but that was immediately orphanned
by the premature death of its discoverer, Karl Schwarszschild,
after which its embarrassing physical implications were hardly taken 
seriously by anyone -- with the notable exception of Oppenheimer 
\cite{Oppy39} -- until the topic was taken up by a less inhibited 
generation  subsequent to the death of Einstein himself, just half a 
century ago, at Princeton in 1955. It was only then (and there)
that John Wheeler inaugurated 
the systematic development of the subject -- for which he coined the 
name ``black hole'' theory -- in a series of pionnering investigations 
that started \cite{Regge57} by addressing the crucial question of 
stability, while not long afterwards, on the other side of the ``iron 
curtain'' another nuclear arms veteran, Yacob Zel'dovich, initiated an
independent approach \cite{Zel66} to the same problem 
(using the alternative name ``frozen star'' which in the end
did not catch on).

\bigskip
{\bf 2. Outcome of stellar evolution:
Chandra's unwelcome limit.}
\medskip

The question of gravitational trapping of light had been raised in the
eighteenth century by Michel and Laplace, whose critical mass 
{\bb \mm\approx \rrho^{-1/2}\fb} assumed the standard mass density that 
is  understood (on the basis of quantum theory as developped by 1930) 
to result in hadronic matter from balance between Fermi repulsion and 
electrostatic attraction which (in Planck units, with proton and 
electron masses {\bb \mm_{\rm p}\approx 10^{-19}\fb}, {\bb \mm_{\rm e}
\approx 10^{-22}\fb}   gives {\bb \rrho\approx\mm_{\rm p}\nn\fb} with 
{\bb\nn\approx\lambda^{-3}\fb} for the Bohr radius {\bb\lambda\approx 
\ee^2/\mm_{\rm e}\fb} with {\bb \ee^2\simeq 1/137\fb}. However most 
theorists refused to face the issue of gravitational collapse even after 
progress in quantum theory lead to Chandrasekhar's 1931 discovery of the 
maximum  mass {\bb \mm\approx \mm_{\rm p}^{-2}\fb} for cold body -- 
which is attained when relativistic gas pressure {\bb \PP\approx
\nn/\lambda\approx\nn^{4/3}\fb} provides the support required by virial 
condition {\bb \PP\approx\mm_{\rm p}^{2/3}\rrho^{4/3}\fb}.
\begin{figure}
\centering
\epsfig{figure=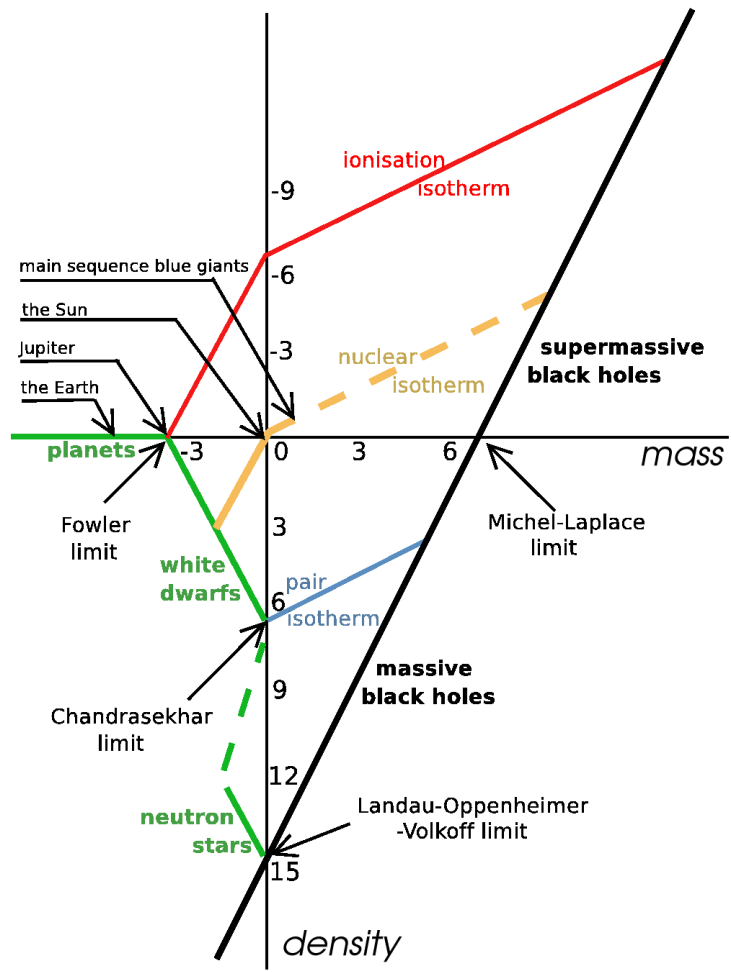, width=9.6 cm}
\caption{Logarithmic plot of density versus mass}
\label{fig001}
\end{figure}

For a lower mass  {\bb \mm\lta \mm_{\rm p}^{-2}\fb}, stellar evolution 
at finite temperature {\bb \Tthet ,\fb} with gas pressure {\bb \PP\approx
\nn \Tthet\fb} subject to {\bb \rrho\approx 
\Tthet^3/\mm_{\rm p}^{\, 3}\mm^2\fb}, can terminate in cold equilibrium  
supported by non-relativistic Fermi pressure {\bb \PP\approx
\nn^{5/3}/\mm_{\rm e}\fb} giving {\bb \rrho\approx 
\mm_{\rm e}^{\,3}\mm_{\rm p}^{\,5}\mm^2\fb} for a white dwarf, or 
{\bb \PP\approx\nn^{5/3}/\mm_{\rm p}\fb} giving {\bb \rrho\approx
\mm_{\rm p}^{\,8}\mm^2 \fb} for neutron star, as shown in Figure 
\ref{fig001}.

However a self gravitating mass of hot gas will be radiation dominated 
with {\fb \PP\approx\Tthet^4\fb},  whenever its mass exceeds the 
Chandrasekhar limit, {\bb \mm\gta \mm_{\rm p}^{-2} ,\fb} so that, 
as first understood by Chandra's Cambridge research director, 
Arthur Eddington, its condition for (thermally supported) equilibrium 
will be given by  {\bb \rrho\approx \Tthet^3/\mm^{1/2} .\fb} What Chandra 
could never get Eddington to accept is that, for such a large mass, no 
cold equilibrium state will be available, so after exhaustion of fuel for 
thermonuclear burning 
(at {\bb \Tthet\approx \ee^4\mm_{\rm p}\fb}) gravitational collapse will 
become inevitable.

\bigskip
{\bf 3. Spherical collapse past the horizon}
\medskip

\begin{figure}
\centering
\epsfig{figure=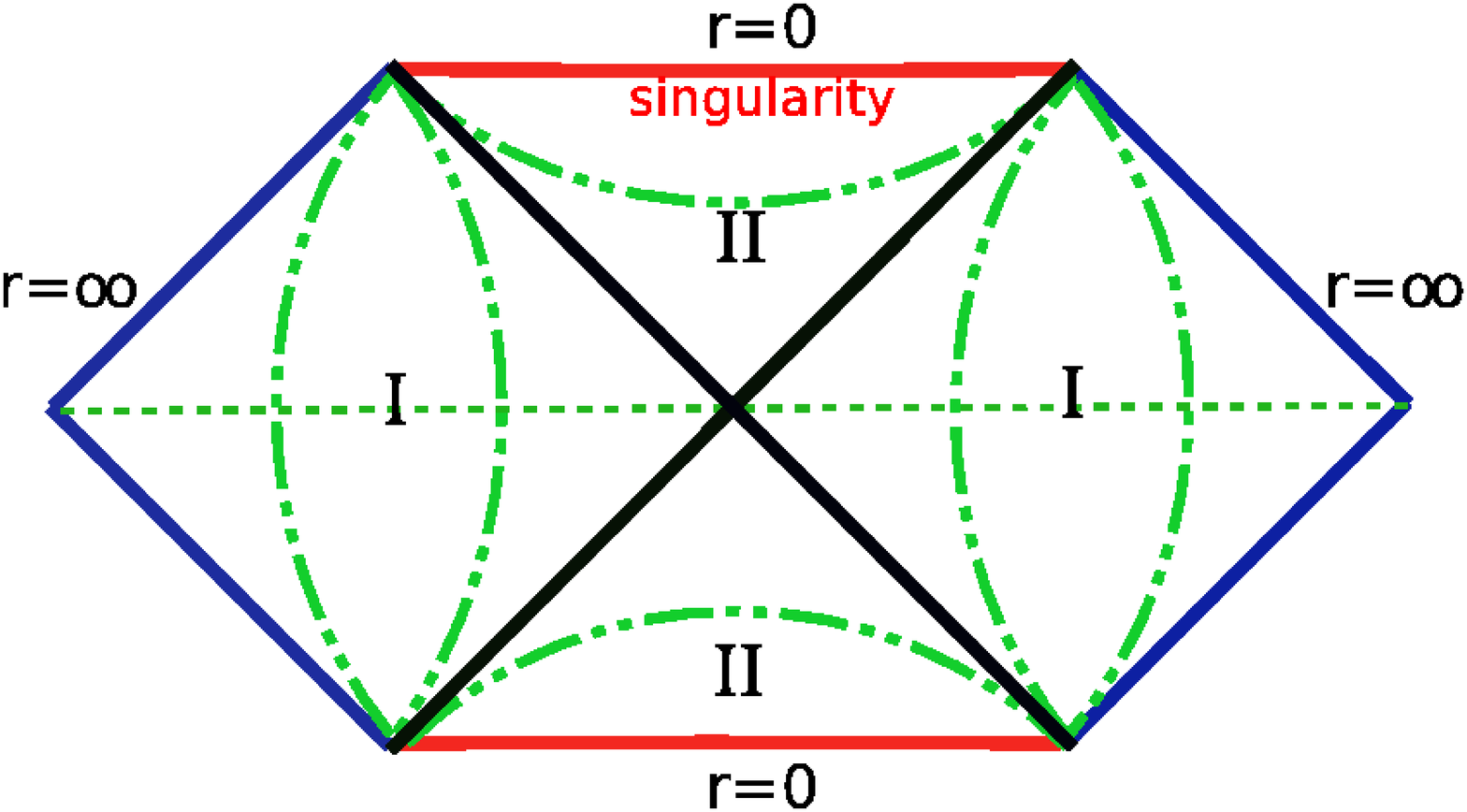, width=7.2 cm}
\epsfig{figure=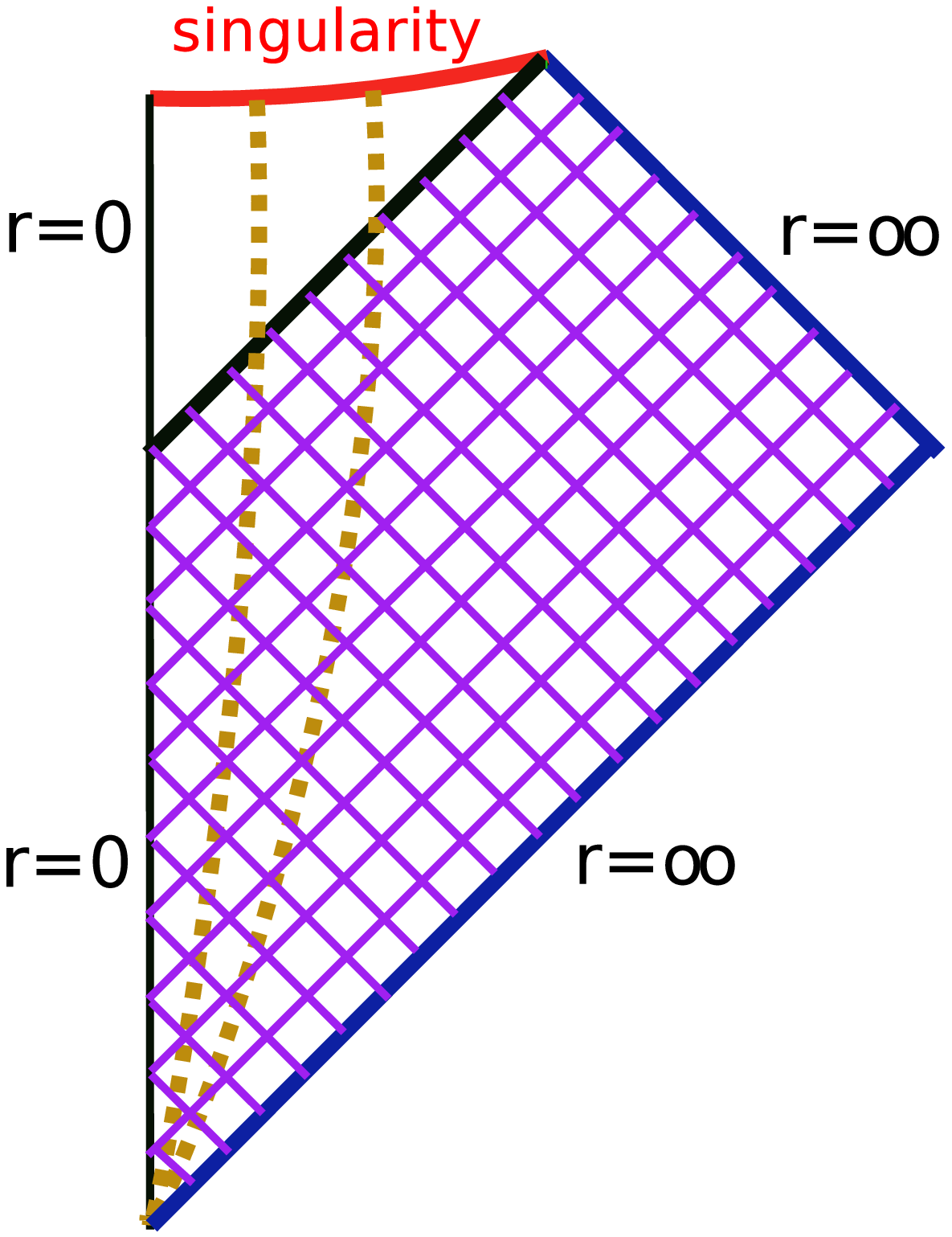, width=3 cm}
\caption{Conformal Projection diagrams showing firstly the combined black 
and white hole geometry obtained (belatedly \cite{Kruskal60} in 1960,
by  Wheeler in collaboration with Kruskal) as the artificial analytic 
(vacuum) extension of the Schwarzschild solution, and secondly a more 
astrophysically natural extension with homogeneous interior (found in 1939 
\cite {Oppy39} by  Oppenheimer in collaboration with Sneyder) in which the 
shaded egion is the ``domain of outer communications'' and the unshaded
region is the prototype example of a hole qualifiable as {\bf black} in 
the strict sense.}
\label{fig002}
\end{figure}

Eddington's example shows how, as has described in detail by Werner 
Israel  \cite{Israel87} (and in striking contrast with the open mindedness 
of Michel and Laplace a century and a half earlier) physicists of 
Einstein's generation tried to convince themselves that nature would never 
allow compacification within a radius comparable to the Schwarzschild 
value. While Einstein  lived, even after Chandrasekhar's discovery had 
shown that such a fate might often be difficult to avoid, the implications 
were taken seriously only by Oppenheimer and his colleagues, who showed 
\cite{Oppy39} how, as shown in Figure \ref{fig002},  the solutions of 
Schwarzschild and Friedmann could be combined to provide a complete 
description of the collapse of a homogeneous spherical body through what 
is now called its event horizon all the way to a terminal singularity. 

Despite the persuasion of such experienced physicists as Wheeler 
and Zel'dovich, and the mathematical progress due to younger
geometers such as Robert Boyer and particularly Roger Penrose, the 
astrophysical relevance of the region near and within the horizon
continued to be widely disbelieved until (and even after)  the 1967 
discovery \cite{Israel67} by Israel of the uniqueness of the 
Schwarzschild geometry as a static solution: many people (for a while
including Israel himself \cite{Israel87}) still supposed (wrongly) that 
the horizon was an unstable artefact of exact spherical symmetry. 
It is therefore not surprising that the question of what such a black 
hole would actually look like was not addressed until much more recently, 
particularly considering that nothing would be seen at all without some 
source of illumination. 
\begin{figure}
\centering
\epsfig{figure=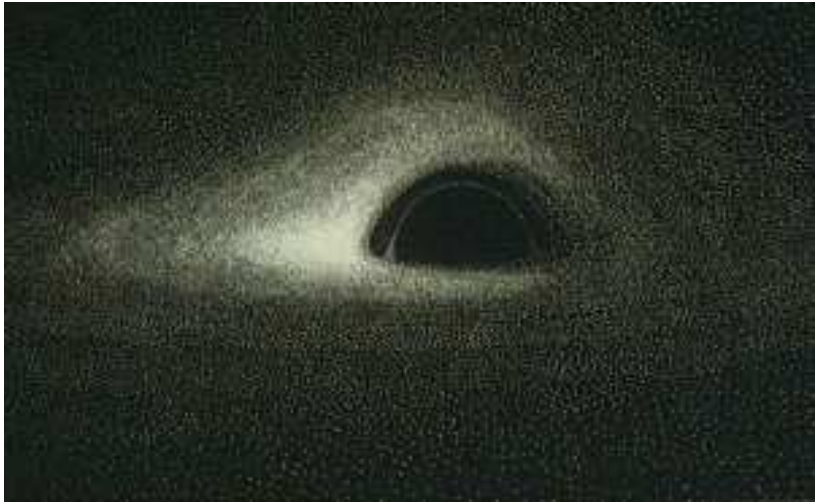, width=10.8cm}
\caption{First  realistic simulation
of distant view of spherical black hole with
thin accretion disc  by Jean-Pierre Luminet, 1978.}
\label{fig0B}
\end{figure} 
\begin{figure}
\centering
\epsfig{figure=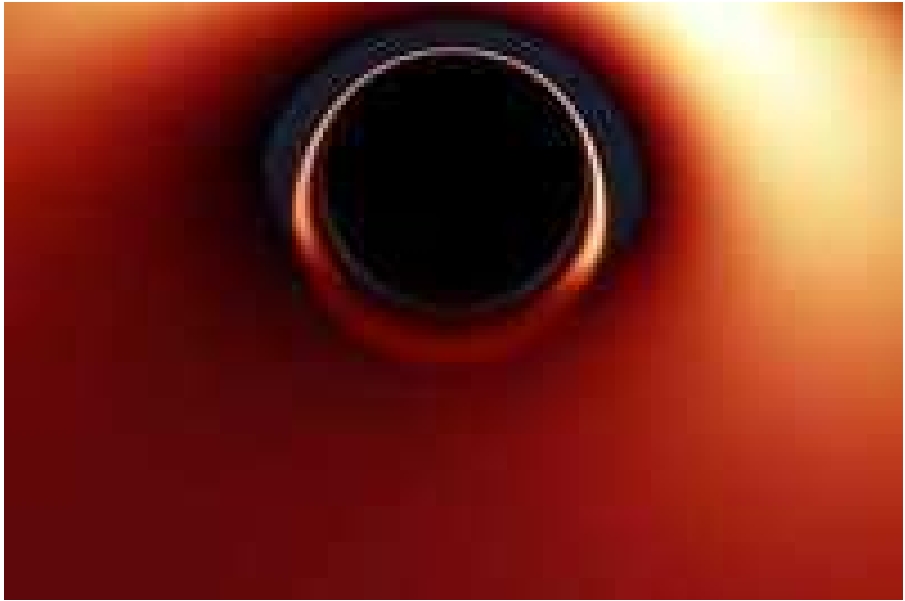, width=12cm}
\caption{Simulation of close up view of spherical black
hole with thin accretion disc by Jean Alain Marck, 1996.}
\label{fig09}
\end{figure}

The realisation that many spectacular astrophysical phenomena ranging 
in scale from supermassive quasars in distant parts of the universe 
down to stellar mass X ray sources within our own galaxy may be 
attributed to accretion discs \cite{Bardeen73,Novikov73,Page74} 
round more or less massive black holes has however provided the motivation 
for increasingly realistic numerical simulations (Figures \ref{fig0B} and 
\ref{fig09}) of what would be seen from outside in the presence of an 
illuminating source of this kind \cite{Luminet79,Marck96}. 

As the most easily calculable example, I have shown in the
appendix how to work out the case shown in Figure \ref{whole} of 
an isolated spherical black hole for which the only source of 
illumination is a uniform distant sky background, viewed as a 
function of proper time, 
{\be\tau=-\frac{4\mm}{3}\left(
\frac{\rr}{2\mm}\right)^{3/2}\, , \label{propertau}\fe}
by  a (doomed)  observer falling towards the  singularity inside 
the black hole, with zero energy and angular momentum. 
\begin{figure}
\centering
\epsfig{figure=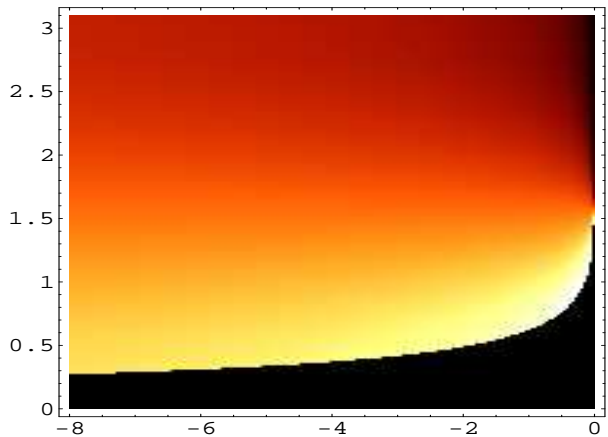, width=8.0 cm, height=6cm}
\caption{\color{navy}
Plot of reception angle {\bb \alph\fb} against
countdown proper time {\bb\tau\fb} for arrival at  singularity of
radially falling zero-energy observer in units such that 
{\bb 2\mm=\sqrt{2/3}\fb} so that null orbit radius {\bb\rr=3\mm\fb}  is 
crossed when {\bb\tau=-1 .\fb} Constant brightness contours indicate 
intensity of light received from uniform background, which will be 
inversely proportional to 4th power of redshift  factor {\bb(1+Z)\fb},
having  unshifted value for rays arriving at 
angle {\bb\pi/2\fb}. 
}
\label{tfig}
\end{figure}
In such a case the redshift {\bb Z\fb} determining the observed energy 
{\bb \calE/(1+Z)\fb} of a photon emitted from the sky background  with 
the uniform average energy $\calE$ say  will be given by the formula
{\be Z=-\frac {{\rm cos} \,\alph}{\sqrt{\rr/2\mm}} 
\, ,\hskip 1 cm \alph>\bet\, ,\label{redshift}\fe}
where {\bb  \alph \fb} is the apparent angle of reception, which must 
of course excede the apparent angle {\bb \bet\fb} subtended by the 
black hole. This means that the redshift will be positive (so that
the sky will appear darker than normal) due to the Doppler effect, for 
photons coming in from behind the observer (with {\bb \alph>\pi/2\fb}). 
However photons received in the range {\bb \bet<\alph<\pi/2\fb} will be 
blueshifted by an amount that will diverge, as shown in Figure 
\ref{tfig}, as the singularity is approached.

\bigskip
{\bf 4. Discovery of horizon stability and of Kerr solution}
\medskip

Following the demise of Einstein (and the development of nuclear 
weapons) a new (less inhibited) generation of physicists, lead by 
Wheeler and Zel'dovich, came to recognise the likelihood -- and need in 
any case for testing -- of stability with respect to non-spherical 
perturbations of what was termed a ``black hole''. Work by 
Vishweshwara \cite{Vishu70}, Price \cite{Price72}, and others confirmed 
that ``anything that can be radiated away will be radiated away'' 
-- leaving a final equilibrium state characterised only by mass and 
angular momentum. The (still open) mathematical question of the extent 
to which this remains true (with singularities hidden inside horizon) 
for very large deviations from sphericity was raised by the ``cosmic 
censorship'' conjecture formulated by Roger Penrose 
\cite{Penrose65,Penrose69} but in any case the relevance of black holes 
for astrophysical phenomena  (notably quasars) was generally accepted 
in astronomical circles from 1970 onwards.

\begin{figure}
\centering
\epsfig{figure=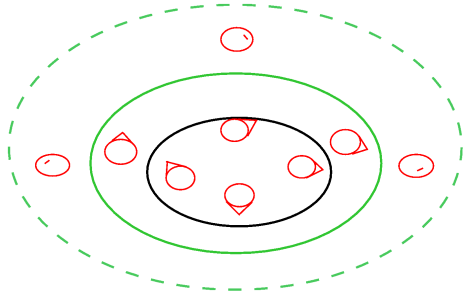, width=6.6 cm}
\caption
{Sketch showing projections of null cones in equatorial space section
of Kerr metric. The pale curve marks the ``ergosurface'' bounding 
the region where stationary motion is allowed, and the heavy black curve 
marks the horizon bounding the trapped ``black hole'' region.}
\label{fig005}
\end{figure}

The generic form of what was afterwards recognised to be the final 
black hole equilibrium state state in question was discovered in 1963, 
when Roy Kerr announced \cite{Kerr63,Kerr65} that ``among the 
solutions ... there is one which is stationary ... and also 
axisymmetric. Like the Schwarzschild metric, which it contains, it is 
type D ... {\bb \mm\fb} is a real constant  ... The metric is
$$ {\rm d}s^2= (\rr^2+\aa^2 {\rm cos}^2\thet)({\rm d}\thet^2
+ {\rm sin}^2\thet\, {\rm d}\pphi^2) +2({\rm d}\uu+\aa\,{\rm sin}^2
\thet\, {\rm d}\pphi)({\rm d}\rr+\aa\,{\rm sin}^2\thet\, {\rm d}\pphi)
$${\be-\!\left(1-\!\frac{2\mm\rr}{\rr^2\!+\!\aa^2{\rm cos}^2\thet}
\right)\!({\rm d}\uu+\aa\,{\rm sin}^2\thet\,{\rm d}\pphi)^2
\, ,\label{Kerrs}\fe}
where {\bb \aa\fb} is a real constant. This may be transformed to an
asymptotically flat coordinate system  ... we find that {\bb \mm\fb} is
the Schwarzschild mass and {\bb \mm\aa\fb} the angular momentum ''.

Since the black hole concept had still not been clearly formulated 
then, it was at first (wrongly) supposed that the physical relevance 
of this vacuum solution would be as the exterior to a compact self
gravitating body like a neutron star, as suggested by Kerr's (off the
mark) conclusion \cite{Kerr63} that it would be ``desirable to 
calculate an interior solution.'' 

\begin{figure}
\centering
\epsfig{figure=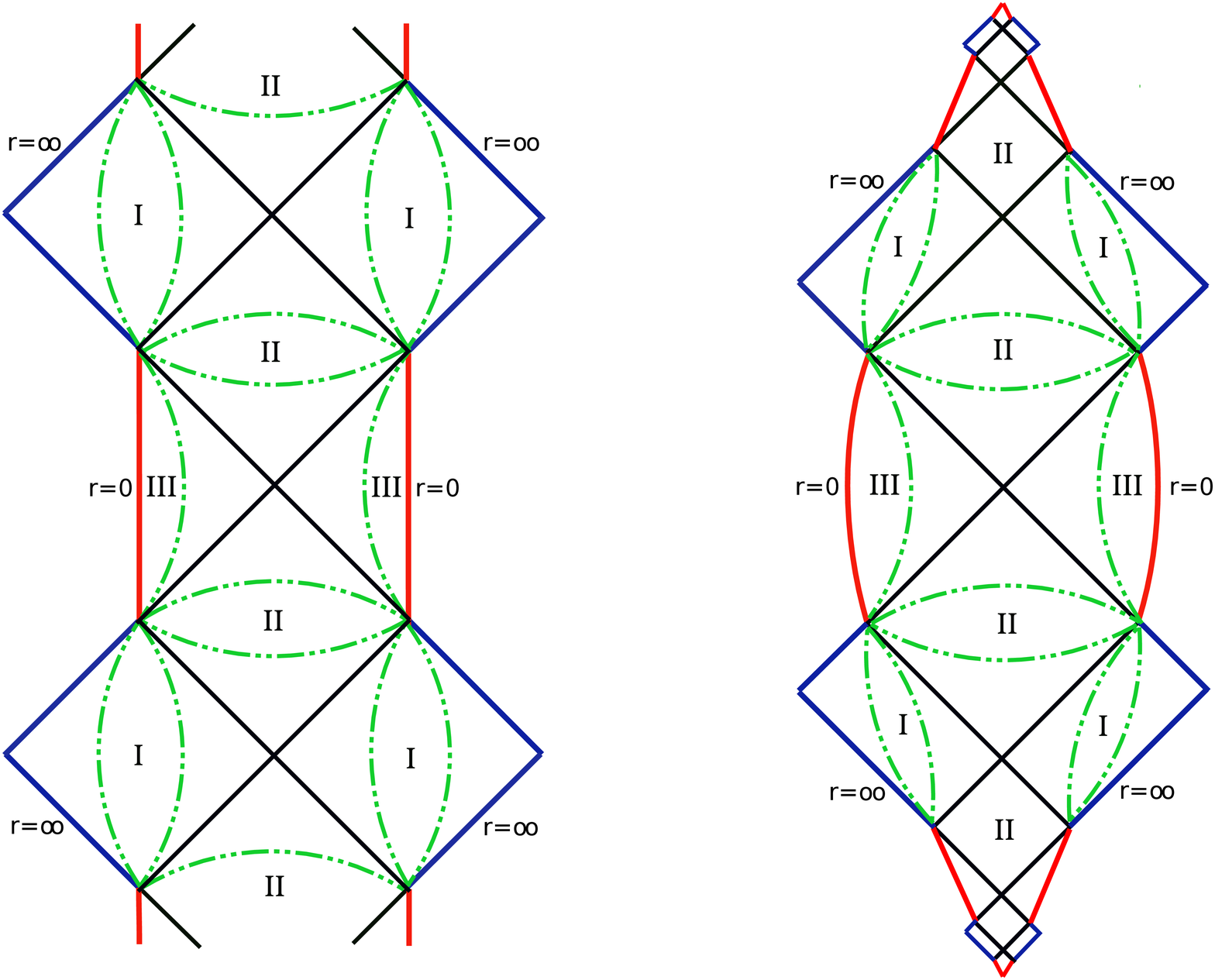, height=6 cm}
\caption
{Representation of equatorial space-time section of Kerr metric
by Conformal Projection diagrams. (Such C.P. diagrams were originally
developed for this purpose.) The second version
achieves complete compactifiction by letting scale for successive
universes tend to zero at extremities of chain.}
\label{fig006}
\end{figure}

What actually makes the Kerr metric so important however, as can be see 
from  Figure \ref{fig006} (using C.P. diagrams, which were originally 
developed for this purpose)  is the feature first clearly  recognised 
\cite{Boyer65,Boyer69}  by Bob Boyer in 1965, which is that for 
{\bb \aa^2\leq \mm^2\fb} the distant sky limit known as ``asymptopia'' is 
both visible and accessible only in a non-singular ``domain of outer 
communications'' bounded by past and future null (outer) {\bf horizons}, 
on which {\bb\Delt=0 ,\fb} where 
{\be \rr =\mm+\cc \, ,\hskip 1 cm \cc=\sqrt{\mm^2-\aa^2}\fe}. 

The topology within the black (and white) hole
 regions was first elucidated \cite{Carter66} in terms of Conformal 
Projections on the symmetry axis in 1966 and  then completely 
\cite{Boyer67,Carter68a,Carter72} by Boyer, Lindquist and myself in 
1967 and 1968 -- the year when the much needed term ``black hole'' was 
finally introduced by Wheeler to describe the region from which light
cannot escape to ``asymptopia''. (A ``white hole'' region would be one 
that could not receive light from``asymptopia''.) In the generic 
rotating case (unlike the static Schwarzschild limit) the well behaved 
domain outside the black hole horizon includes an ``ergosphere'' region
where, as shown in Figure \ref{fig005}, the Killing vector generating 
the stationarity symmetry becomes spacelike, so that (globally
defined) particle energies can be negative. 

\begin{figure}
\centering
\epsfig{figure=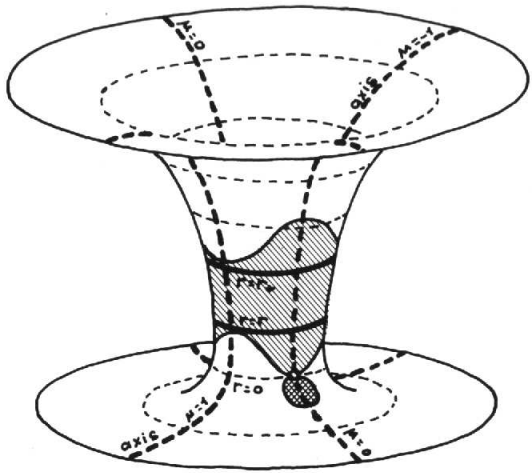, height=6.6 cm}
\caption{Reproduction from 1972 Les Houches notes \cite{Carter72}:
sketch of  {\bb \{\rr,\thet\}\fb} section through ring 
singularity at junction between  heavily shaded region responsible for 
causality violation where axisymmetry generator is spacelike, and lightly
 shaded ergo region where time summetry generator is spacelike.
}
\label{fig1}
\end{figure}

In contrast with the good behavior of the outer region,  
{\bb \rr>\mm+\cc ,\fb} I  found that,  as well as having the irremovable 
ring shaped  curvature singularity  already noticed by Kerr where 
{\bb \rr^2\!+\!\aa^2 {\rm cos}^2\thet\!\rightarrow\! 0 ,\fb}, the inner 
parts of the rotating Kerr solutions would always be causally 
pathological, due to the existence near the ring singularity  
of a small region (see Figure \ref{fig1}) where the axial symmetry 
generating Killing vector becomes timelike\cite{Carter68a, Carter78}.
This feature gives rise to a causality violating
``time machine region'' (a feature so ``outrageous'' as to be 
unmentionable even by Thorne \cite{Thorne94})  that would extend 
all the way out to ``asymptopia'' (meaning {\bb \rr\rightarrow 
+\infty\fb}) in the -- presumably unphysical -- case for which 
{\bb \aa^2>\mm^2\fb}. (I would emphasize that this kind of time 
machine, like those recently considered by Ori \cite{Ori05}, would 
survive even if one takes the covering space, unlike a time machine of 
the wormhole kind discussed by Thorne \cite{Thorne94} which is merely 
an artefact of  multiply connected space time topology).

In so far as the (physically relevant) black hole cases characterised 
by {\bb \aa^2\leq\mm^2\fb} are concerned, the good news 
\cite{Carter68a, Carter78} (for believers in causality) is that the 
closed timelike lines are all contained within the inner region 
{\bb \rr<\mm-\cc  .\fb} The boundary of the time machine region is 
constituted by the ``inner horizon'', where  {\bb \rr=\mm-\cc  .\fb} 
which acts as a Cauchy hypersuface from the point of view of inital
data for formation of the black hole by gravitational collapse.
Unlike the outer horizon {\bb \rr=\mm+\cc  ,\fb} whose stability
throughout the allowed range {\bb 0\leq \aa^2<\mm^2\fb} has been
even confirmed by Whiting \cite{Whiting89}, it was to 
be expected \cite{Simpson73, Hartle82} that a Cauchy horizon
of the kind occurring at  {\bb \rr=\mm-\cc  \fb} would
be unstable, and it has been shown that 
outcome is likely to be the formation of a curvature singularity
of the weak kind designated by the term ``mass inflation''
\cite{Poisson90,Ori92,Ori96}.

\bigskip
{\bf 5. Seductive mathematical features of Kerr type metrics}
\medskip

In his original 1963 letter \cite{Kerr63}, and with Alfred Schild 
\cite{Kerr65} in a sequel, Kerr obtained the useful alternative form
{\be {\rm d} s^2=g_{\mu\nu}\,{\rm d}x^\mu{\rm d}x^\nu
=\eta_{\mu\nu}+2(\mm/\UU)\lle_\mu\lle_\nu\fe} 
 with  {\bf null} covector
{\bb \lle_\mu {\rm d}x^\mu\!={\rm d}\uu+\aa\,{\rm sin}^2\thet
\, {\rm d}\pphi\, ,\fb}
for
{\bb \UU=(\rr^2+\aa^2 {\rm cos}^2\thet)/\rr\, ,\fb}
in a {\bf flat} background. The latter was obtained in
the Minkowski form,
{\be \eta_{\mu\nu}\,{\rm d}x^\mu{\rm d}x^\nu={\rm d}\bar \xx^2\!+\!{\rm d}
\bar \yy^2\!+\!
{\rm d}\bar \zz^2\!-{\rm d}\bar \tt^2\, ,\fe}
by setting
{\bb \bar\tt=\uu-\rr\, ,\fb} {\bb \bar \zz=\aa\,{\rm cos}\,\thet\, ,\fb} 
{\bb\bar \xx+i \bar \yy=(\rr-i\aa)
{\rm e}^{i\pphi}{\rm sin}\,\thet\, ,\fb} which gave
{\be \lle_\mu {\rm d}x^\mu\!= {\rm d}\bar \tt+\!
\frac{ \bar \zz {\rm d} \bar \zz}{\rr}\!+\frac
{(\rr \bar \xx\!-\!\aa \bar \yy){\rm d}\bar \xx\!+\!(\rr \bar \yy\!
+\!\aa \bar \xx){\rm d} \bar \yy}{ \rr^2+\aa^2}\, .\fe}
(This form of pure vacuum solution was generalised to higher dimensions  
by Myers and Perry\cite{Myers86}. It is perhaps of greater current
cosmological interest -- in view of the evidence that the expansion of 
the universe is accelerating -- that this form has also beeen extended 
to include a cosmological constant in a 4 dimensional {\bf De Sitter} 
background by myself \cite{Carter72, Gibbons77}, while further 
generalisations to a De Sitter background in 5 and higher dimensions 
\cite{Hawking98,Gibbons04} have been obtained more recently.)

As well as time and axial symmmetry, the Kerr solution has a discrete 
PT symmetry that was predictable from Papapetrou's ``circularity'' 
theorem \cite{Papapetrou66}, and made manifest in 1967 \cite{Boyer67} 
by the Boyer Lindquist transformation
{\be {\rm d}\tt={\rm d}\uu-(\rr^2+\aa^2)\Delt^{-1}{\rm d}\rr\, ,\ \
{\rm d}\phhi=-{\rm d}\pphi+\aa\,\Delt^{-1}{\rm d}\rr \, ,\fe}
with  
{\be \Delt=\rr^2-2\mm\rr+\aa^2 \, .\fe}
This gives Kerr's null form as
{\be  \lle_\mu {\rm d}x^\mu\!=
 {\rm d}\tt-\aa\,{\rm sin}^2\thet\,{\rm d}\phhi+\varho^2
\Delt^{-1}{\rm d}\rr\, ,\hskip 1 cm
\varho\!=\!\sqrt{\!\rr^2\!+\!\aa^2{\rm cos}^2\thet} , .\fe}
The metric itself is thereby obtained in the convenient form
{\be  {\rm d}s^2= \varho^2\left(\frac{{\rm d}\rr^2}{\Delt}+
{\rm d}\thet^2 \right)+(\rr^2+\aa^2){\rm sin}^2\thet\,{\rm d}\phhi^2 
  +\frac{2\mm\rr}{\varho^2}({\rm d}\tt-\aa\,{\rm sin}^2\thet
\,{\rm d}\phhi)^2\!-{\rm d}\tt^2\, ,\fe}
in which there are cross terms involving the non-ignorable 
differentials, {\bb{\rm d}\rr\fb} and {\fb {\rm d}\thet\, ,\fb} but 
-- as the price for this simplification -- if {\bb \aa^2\leq\mm^2\fb}
there will be a removable coordinate singularity on the null 
``horizon'' where {\bb \Delt\fb} vanishes.

Whereas the possibility of making the foregoing simplification was 
predictable in advance, there was no reason to anticipate the
discovery \cite{Carter68a,Carter68b} that, in addition to the ordinary 
``circular'' symmetry  generated by Killing vectors, {\bb
\kk^\mu\partial/\partial x^\mu= \partial/\partial \tt\fb} and
{\bb\hh^\mu\partial/\partial x^\mu= \partial/\partial \phhi\, ,\fb}
the Kerr metric would turn out to have the hidden symmetry that is
embodied in the {\bf canonical} tetrad 
{\be g_{\mu\nu}\!=\sum_{_{i=1}}^{\,_3}
\varth^{\hat{i}}_{\,\mu}\varth^{\hat{i}}_{\nu}
-\!\varth^{\hat{^0}}_{\mu}\varth^{\hat{^0}}_{\nu}\!\fe} 
specified by
{\be \varth^{\hat{^1}}_{\mu}{\rm d}x^\mu\!
=\!(\!\varho/\sqrt{\!\Delt}){\rm d}\rr \, ,\hskip 1cm
\varth^{\hat{^2}}_{\mu}{\rm d}x^\mu\!
=\varho\,{\rm d}\thet ,\fe}
{\be\frac{\varth^{\hat{^3}}_{\mu}{\rm d}x^\mu\!}{{\rm sin}\,\thet}
=\frac{(\rr^2\!+\!\aa^2) {\rm d}\phhi\!-\!\aa {\rm d}\tt}{\varho}\, , 
\ \ \frac{\varth^{\hat{^0}}_{\mu}{\rm d}x^\mu\!}{\sqrt\Delt}=\frac
{{\rm d}\tt\!-\!\aa\,{\rm sin}^2\thet\,{\rm d}\phhi}{\varho}
\, .\fe}

In terms of this canonical tetrad, the Kerr-Schild form of the metric
is expressible as 
{\be g_{\mu\nu}=\eta_{\mu\nu}\!+\!2\mm\rr(\varth^{\hat{^0}}_{\mu}\!+\!
\varth^{\hat{^1}}_{\mu})(\varth^{\hat{^0}}_{\nu}\!+
\!\varth^{\hat{^1}}_{\nu})\, ,\fe}
while the Killing-Yano 2-form brought to light by Roger Penrose
and his coworkers is expressible as
{\be \ff_{\mu\nu} =2\aa\, {\rm cos}\,\thet\,
\varth^{\hat{^1}}_{[\mu}\varth^{\hat{^0}}_{\nu]}-2\rr \,
\varth^{\hat{^2}}_{[\mu}\varth^{\hat{^3}}_{\nu]}\, .\fe}
The property of being a Killing-Yano 2-form means that it is such as to 
satisfy the very restrictive condition condition 
{\be \nabla_{\!\mu}\ff_{\nu\rho}=
\nabla_{\![\mu}\ff_{\nu\rho]}\, ,\fe}
thus providing a symmetric solution
{\be \KK_{\mu\nu}=\ff_{\mu\rho}\ff^\rho_{\ \nu}\, ,\hskip 1 cm
 \nabla_{\!(\mu}\KK_{\nu\rho)}=0\, ,\fe}
of the Eisenhart type Killing tensor equation,
as well as secondary and primary solutions
{\bb \tilde \kk{^\mu}=\KK^\mu_{\ \nu}\kk^\nu
=\!\aa^2\kk^\mu\!+\!\aa\hh^\mu\fb} and
{\bb \kk^\mu =\frac{_1}{^6}\varepsilon^{\mu\nu\rho\sigma}
\nabla_{\!\nu}\ff_{\rho\sigma}\fb} of the ordinary
Killing vector equation {\bb \nabla_{\!(\mu}\kk_{\nu)}=0 \, .\fb}

For affine geodesic motion,  {\bb  \pp^\nu\nabla_{\!\nu} \pp^\mu\!
=\! 0 ,\fb} one thus obtains (energy and axial angular momentum) 
constants {\bb \calE\!=\!\kk^\nu
\pp_\nu \fb} and {\bb \calM=\!\hh^\nu \pp_\nu ,\fb}
while the Killing tensor gives the constant
 {\bb \calK=\KK^{\mu\nu}\pp_\mu\pp_\nu =
\elle_\mu\elle^\nu\ ,\fb}  with (angular momentum)
{\bb \calJ_\mu=\ff_{\mu\nu} \pp^\mu\fb}  obeying
{\bb \pp^\nu\nabla_{\!\nu}\elle_\mu=0 .\fb}

There will also \cite{Carter77} be corresponding (self adjoint) 
operators 
{\be \calE\!=\!i\kk^\nu\nabla_{\!\nu} \, ,\hskip 1 cm 
\calM\!=\!i\hh^\nu\nabla_{\!\nu} \, ,\hskip 1 cm 
\calK=\nabla_{\!\mu} \KK^{\mu\nu}\nabla_{\!\nu}\, ,\fe} 
whose action on a scalar field commute with that of the
the Dalembertian {\bb \square=\nabla^\nu\nabla_{\!\nu}\fb}: 
in other words {\bb [\calE,\square]=0 ,\fb} {\bb [\calM,\square]
=0 ,\fb} and (consistently with the integrability condition 
{\bb \KK^\rho{_{[\mu}}R_{\nu]\rho}=0\fb}) also  
{\bb [\calK,\square]=0\fb}.

The ensuing integrability of the geodesic equation \cite{Carter68a}
and of the scalar wave equation is equivalent to their solubility
by separation of variables\cite{Carter68a,Carter68b}. The possibility
of extending these rather miraculous separability properties to the
neutrino equation \cite{Unruh73} and even to the massive spin 1/2 
field \cite{Chandra76,Page76} as governed by the Dirac operator 
{\bb \calD=\gamma^\mu\nabla_{\!\mu}\fb} is attributable to 
corresponding spinor operator conservation laws 
 {\be [\calE,\calD]=0 \, ,\hskip 1 cm [\calM,\calD]=0\, ,
\hskip 1 cm[\calJ,\calD]=0\, ,\fe}
of energy, axial angular momentum,   and (unsquared) total angular  
momentum, as respectively given  \cite{Carter79} by
{\be\calE\!=\!i\kk^\nu\nabla_{\!\nu}\!+\!\frac{_1}{^4}i(\nabla_{\!\mu}
\kk_\nu)\gamma^\mu\gamma^\nu\, ,\hskip 1 cm  \calM\!=
\!i\hh^\nu\nabla_{\!\nu}\!+\!\frac{_1}{^4}i(\nabla_{\!\mu}
\hh_\nu)\gamma^\mu\gamma^\nu \, ,\fe} and 
{\be\calJ=i\gamma^\mu(\gamma^{_5}\ff_{\!\mu}^{\  \nu}
\nabla_{\!\nu}-\kk_\mu)\, .\fe}

Such a neat commutation formulation is not (yet?) available for 
Teukolsky's extension \cite{Teukolsky73,Press73} of solubility by 
separation of variables to massless spin 1 and spin 2 fields 
representing electromagnetic and gravitation perturbations -- of 
which the latter are particularly important for Bernard Whiting's  
demonstration \cite{Whiting89} of stability. An even more difficult 
problem is posed by the charged generalisation \cite{Newman65} of 
the Kerr black hole metric, which retains many of its convenient
properties (and is noteworthy for having the same gyromagnetic ratio
as the Dirac electron \cite{Carter68a,Treves75}) but which gives rise 
to a system of coupled electromagnetic and gravitational perturbations 
that has so far been found to be entirely intractible.

\bigskip
{\bf 6. No hair and uniqueness theorems for black hole equilibrium}
\medskip

The overwhelming importance of Kerr solution derives from its provision 
of  the generic representation of the final  outcome of gravitational
collapse, as was made fairly clear in 1971  by the prototype {\bf no-hair 
theorem} \cite{Carter71,Carter92} proving that no other vacuum black hole 
equilibrium state can be obtained by continuous axisymmetric variation 
from the spherical Schwarzschild solution that had been shown by the 
earlier work of Israel \cite{Israel67} (before the generic definition of 
a black hole was available) to be only static possibility.

Conceivable loopholes (such as doubts about the axisymmetry assumption) 
in the reasonning leading to this conclusion (which was rapidly -- 
perhaps too uncritically -- accepted in astronomical circles) were 
successively dealt with by the subsequent mathematical work of Stephen 
Hawking \cite{Hawking72}, David Robinson \cite{Robinson75} and other 
more recent contributors \cite{Sudarsky93,Chrusciel93,Chrusciel94} to 
what has by now become a rather complete and watertight uniqueness 
theorem for pure vaccum black hole solutions in 4 spacetime dimensions. 
It should however be remarked \cite{Carter99} that there are some 
mathematical loose ends (concerning assumptions of analyticity and 
causality) that still need to be tidied up.

The demonstration uses ellipsoidal coordinates for the 2-dimensonal
space metric 
{\bb{\rm d}\hat s^2 ={\rm d}\lambda^2/(\llambda^2\!-\!\cc^2)+{\rm d}
\mmu^2/(1\!-\!\mmu^2) ,\fb}
in terms of which the generic stationary axisymmetric asymptotically 
flat vacuum metric is known from the work of Papapetrou 
\cite{Papapetrou66} to be expressible in the form
{\be {\rm d}s^2= \varho^2{\rm d}\hat s^2 +\XX({\rm d} \phhi-
\omeg\, {\rm d}\tt)^2-(\llambda^2\!-\!\cc^2)(1\!-\!\mmu^2)
{\rm d}\tt^2 \, .\fe}
for which, by the introduction of  an Ernst \cite{Ernst68} type potential
given by {\bb \XX^2\partial \omeg/\partial \llambda = 
 (1\!-\!\mmu^2) \partial \YY/\partial\mmu  ,\fb}  the relevant Einstein 
equations will be obtainable from the  (positive definite) action
{\bb \int {\rm d}\llambda\,{\rm d}\mmu\,
(|\hat\nabla \XX|^2+|\hat\nabla \YY|^2)/\XX^2 .\fb}

The black hole equilibrium problem is thus 
\cite{Carter71,Carter72,Carter92} reduced to a non
linear 2 dimensional elliptic boundary value problem for the
scalars {\bb \XX, \YY ,\fb} subject to
conditions of regularity on the horizon (with rigid angular  
velocity {\bb\Ome\fb}) where {\bb \llambda=\cc\fb}
and to appropriate boundary conditions 
on the axis where {\bb \mmu=\pm 1\fb} and at large radius {\bb \llambda
\rightarrow \infty\fb} in terms of angular momentum {\bb\mm\aa\, .\fb}

The uniqueness theorem states that this 2 dimensional boundary problem  
has  no solutions other that those given given (with 
{\bb\llambda=\rr-\mm ,\fb} { \bb \mmu ={\rm cos}\, \thet)\fb} by the 
Kerr solution having mass {\bb \mm=\sqrt{\cc^2+\aa^2}\fb} and horizon 
angular velocity {\bb\Ome=\aa/2\mm(\mm+\cc)\ .\fb} The proof is obtained 
from an identity equating a quantity that is a positive definite 
function of the relevant deviation (of some other hypothetical solution 
from the Kerr value) to a divergence whose surface integral can be seen 
to vanish by the boundary conditions.

The original no-hair theorem (applying just to the small deviation
limit) was based on an infinitesimal divergence identity that I obtained 
by a hit and miss method \cite{Carter71} that was generalised by
Robinson \cite{Robinson75} to the finite difference divergence identity 
that was needed to complete the proof in the pure vacuum case.
For the electromagnetic (Einstein Maxwell) generalisation, the 
analogous step from an infinitesimal no-hair theorem\cite{Robinson74} 
to a fully non-linear uniquenes theorem was more difficult, and was
not obtained until our hit and miss approach was superceded by the
more sophisticated methods that were developed later on by 
Mazur \cite{Mazur82,Mazur84} and Bunting \cite{Bunting83,Carter85}. 

\bigskip
{\bf 6. Further developments}
\medskip

\begin{figure}
\centering
\epsfig{figure=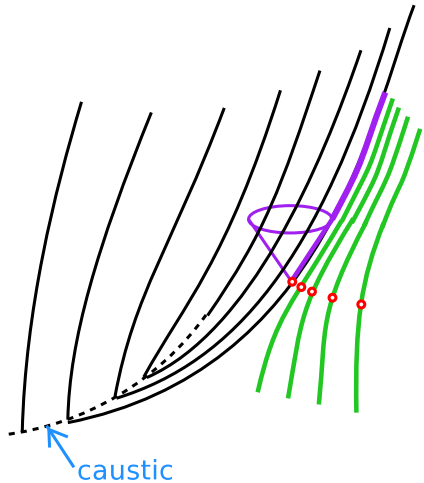, width=5.6 cm}
\caption{Sketch (in plot of time against space) of segment of black 
hole horizon showing how a null generator (obtained as limit of 
escaping timelike curves) can begin (on a caustic) but can never end 
towards the future.}
\label{cusp}
\end{figure}

After it had become clear that (in the framework of Einstein's 
theory) the Kerr solutions (with {\bb \aa\leq \mm \fb}) are the only 
vacuumm black hole equilibrium states, the next thing to be 
investigated was the way the black holes will evolve when the 
equilibrium is perturbed. A particularly noteworthy result, based
on concepts (see Figure \ref{cusp} ) developed in collaboration
with Penrose \cite{Hawking69}  was the demonstration by Stephen
Hawking \cite{Hawking71,Hawking72} that the area of a black hole
horizon (which is proportional to what Christodoulou \cite{Christo70}
had previously identified as irreducible mass) can never decrease. 
More particularly it was shown \cite{Hartle72} that the area would 
grow, not only when the hole swallowed matter but more generally 
whenever the null generators of the horizon were subjected to shear. 
It was remarked that this effect could be described in terms of an 
effective viscosity and that the horizon could also be characterised 
\cite{Damour78, Znajek78} by an effective resistivity. 

Later astrophysical developments were concerned more with surrounding
or infalling matter -- for example  in accretion discs -- than 
with the black hole as such, at least until recently. However the 
prospect of detecting gravitational radiation in the foreseeable 
future has encouraged a resurgence of interest in purely gravitational 
effects, particularly those involved in binary coalescence. The climax 
of a coalescence is too complicated to be dealt with except by advanced 
methods of numerical computation, but the quasi stationary preliminary 
stages are more amenable
\cite{York83,Matzner85,Cook91,Pfeiffer00,Bona02,Cook02},
as also are the final stages of ringdown, which can be analysed in 
terms of quasi normal modes (and their superpositions in power law 
tails of the kind first described by Price \cite{Price72}) which have 
been the subject of considerable attention, particularly concerning the 
influence of rotation \cite{Leaver85,Seidel90,Kokko91,Onoz97,Laguna97}.

\bigskip

{\bf Appendix: null geodesics in spherical case.}
\medskip

Although it would be insufficient for the complete Kruskal (black and 
white hole) extension, in order to cover a purely black hole 
(Oppenheimer Sneyder type) extension of the Schwarzschild solution,  
it will suffice to use an outgoing null coordinate patch of the 
kind introduced for the Kerr metric (\ref{Kerrs}) for which, when 
{\bb \aa=0 , \fb} the metric will be given in terms of 
{\bb x^{_0}=\uu ,\fb} {\bb x^{_1}=\rr ,\fb} {\bb x^{_2}=\thet ,\fb} 
{\bb x^{_3}=\pphi ,\fb}, simply by
{\be {\rm d}s^2=
-\!\left(1-\!{2\mm/\rr}\right) {\rm d}\uu^2
+2\,{\rm d}\uu \,{\rm d}\rr+
\rr^2{\rm d}\thet^2+ {\rm sin}^2\thet\, {\rm d}\pphi^2
\, .\label{S}\fe}

Within such a system, an observer falling in freely from a large 
distance with zero energy and angular momentum will have a geodesic 
trajectory characterised by fixed values of the angle coordinates 
{\bb \thet\fb} and {\bb \pphi\fb} and by a radial  coordinate 
{\bb \rr\fb} that is given implicitly as a monotonically decreasing 
function of proper time by (\ref{propertau}) and as a 
monotonically decreasing function of the ignorable coordinate
{\bb \uu \fb} by the relation 
{\be \frac{\uu_0-\uu}{2\mm}=\frac{2}{3}\sqrt{\frac{\rr}{2\mm}}
\left(\frac{\rr}{2\mm}+3\right)-\frac{\rr}{2\mm}-
2\,{\rm ln}\left\{1+\sqrt{\frac{\rr}{2\mm}}\right\}\, ,\fe}
in which {\bb \uu_0\fb} is a constant of integration specifying
the value of {\bb \uu\fb} for which the trajectory terminates
at the singular limit {\bb\rr\rightarrow 0\fb}.
For such a trajectory the (future oriented)
timelike unit tangent vector will be given by
{\be e_{_{(0)}}^{\ _0}=\left(1+\sqrt{2\mm/\rr}
\right)^{-1}
\, ,\hskip 1 cm e_{_{(0)}}^{\ _1}= -\sqrt{2\mm/\rr}
\, ,\fe}
and the tetrad specifying a corresponding local reference frame can be 
completed in a natural manner by using the associated (outward oriented) 
orthogonal spacelike unit vector, which will be given by
{\be e_{_{(1)}}^{\ _0} =\left(1+\sqrt{2\mm/\rr}
\right)^{-1}
\, ,\hskip 1 cm e_{_{(1)}}^{\ _1}=1\, ,\fe}
together with two other horizontally oriented unit vectors whose 
specification will not matter for the present purpose because of the 
rotation symmetry of the system.

Let us consider the observation of photon that arrives with trajectory 
deviating by an angle {\bb\alph\fb} say from the outward radial direction. 
The first two components of its (null) momentum vector, {\bb \pp^\mu\fb} 
say,  will evidently be given in terms of its energy {\bb\tilde \calE\fb} 
with respect to such a frame by
{\be \pp^{\, i}=\tilde\calE(e_{_{(0)}}^{i}+{\rm cos}\,\alph\,
e_{_{(1)}}^{i})\, ,\fe}
for {\bb i= 0,1\fb}. This locally observed energy is to be compared with
the globally defined photon energy (as calibrated with respect to the 
asympotic rest frame at large distance) that will be given in terms of  
the timelike Killing vector with components 
{\bb \kk^\mu=\delta_{_0}^\mu\fb}  by
{\be \calE=-\kk_\mu\pp^\mu= \left(1-\!{2\mm/ \rr}\right)\pp^{_0}
-\pp^{_1}\, ,\fe}
the important feature of the latter being that it is conserved by the 
affine transport of the momentum vector along the null geodesic photon 
trajectory to which it is tangent.  It can thus be seen that the 
corresponding locally observed energy {\bb\tilde\calE\fb} will be 
related to the globally defined energy constant {\bb \calE\fb} by
{\be \calE=\tilde\calE(1+Z)\, ,\label{shift} \fe}
with the redshift {\bb Z\fb} given by (\ref{redshift}),
and that the associated component ratio will be given by
{\be\frac{\pp^{\,_1}}{\pp^{\,_0}}=(1+\sqrt{2\mm/\rr})\left(\frac
{{\rm cos}\,\alph-\sqrt{2\mm/\rr}}{{\rm cos}\,\alph +1}\right)
\, .\label{ratio}\fe}

As the (unsurprising) spherical limit of the (still rather mysterious)
separability of Kerr's rotating generalisation, the evolution of the
relevant affinely transported momentum components will be given in
terms of the energy constant {\bb \calE\fb} and the associated squared
angular momentum constant {\bb \calK\fb}~\cite{Carter68a} (using a dot
for differentiation with respect to the affine parameter) by
{\be \pp^{\,_0}=\dot \uu=\frac{\pm\sqrt\calR +\calP}{\rr^2-2\mm\rr}\, ,
\hskip 1 cm \pp^{_1}=\dot\rr=\frac{\pm\sqrt\calR}{\rr^2}\, ,\fe}
where
{\be\calP=\calE\rr^2\, ,\hskip 1 cm
\calR=\calP^2-\calK(\rr^2-2\mm\rr)\, .\label{radialf}\fe}
We thereby obtain
{\be\frac{\pp^{\,_0}}{\pp^{\,_1}}=\frac{{\rm d}\uu}{{\rm d}\rr}=
(1-2\mm/\rr)^{-1}
\left( 1\pm\frac{\calP}{\sqrt{\calR}}\right)\, ,\label{radfn}\fe}
and hence, by comparison with (\ref{ratio}),
{\be\pm\frac{\sqrt{\calR}}{\calP}=\sqrt{\rr/2\mm}\left(\frac
{ {\rm cos}\,\alph-\sqrt{2\mm/\rr}}{\sqrt{\rr/2\mm} -{\rm cos}\,
\alph}\right)\label{signdisc}\, .\fe}
This expression can be used to evaluate the squared angular momentum
constant as a function of the locally defined energy
{\bb\tilde\calE\fb} and angle {\bb \alph\fb} in the form
{\be \calK=\rr^2\tilde\calE{^2}{\rm sin}^2\alph\label{angconst}\fe}
in which the variable {\bb\tilde\calE\fb} will itself be given 
via (\ref{shift}) in terms of the globally defined energy constant 
{\bb\calE\fb} by the -- red or blue -- shift formula (\ref{redshift}) 
so that for the square of the constant ratio of angular momentum to 
energy one obtains
{\be \frac{\calK}{\calE^2}=\left(\frac{\rr\,{\rm sin}\,\alph}
{1-\sqrt{2\mm/\rr}\,{\rm cos}\,\alph}\right)^2\, .\label{angrat}\fe}

With respect to an unconventional affine parameter orientation 
condition to the effect that the energy should always be non-negative,
{\bb\calE\geq 0\, ,\fb}  it can be seen (in view of the consideration
that the squared angular momentum constant must necessarily be
non-negative, {\bb\calK\geq 0\fb}) that a null geodesic segment will be
appropriately be describable as ``incoming'' or ``outgoing'' according 
to whether the upper or lower of the sign possibilities {\bb\pm\fb} is
applicable, i.e. according to whether the right hand side of 
(\ref{signdisc}) is positive or strictly negative. It is however to be
 remarked that this convention will be consistent with the usual 
requirement that the affine parameter orientation be future oriented, 
giving {\bb \dot\uu\geq 0\fb}, only outside the horizon and  for 
``ingoing'' null segments within the horizon, where {\bb\rr<2\mm\, ,\fb} 
but that for ``outgoing'' null segments within the horizon it would 
entail the opposite orientation convention, giving
{\bb \dot\uu\leq 0\fb}. With respect to the usual parameter orientation
condition giving {\bb \dot\uu\geq 0\fb} the ``outgoing'' null segments
within the horizon will need to be parametrised the other way round,
which means that they will be characterised by negative energy
{\bb \calE<0\fb} and by the upper of the sign possibilities {\bb\pm\fb}.

Whichever convention is used, it can be seen that within the horizon
the radius {\bb\rr\fb} will aways be a decreasing function of {\bb\uu\fb},
even for the (relatively) ``outgoing'' null segments, and that only
an ``incoming'' null segment can cross the horizon at a
finite value of {\bb\uu\fb}. It can be seen from (\ref{signdisc}) that
outside the horizon (i.e. for {\bb\rr>2\mm\fb})  the criterion for a null
segment to be classified as ``incoming'' is that it should have
{\bb {\rm cos}\,\alph\leq\sqrt{2\mm/\rr}\fb} and that the
corresponding requirement within the horizon is
{\bb {\rm cos}\,\alph\leq\sqrt{\rr/2\mm}\, .\fb}

It can be deduced from the expression (\ref{radialf}) that the function
{\bb\calR\fb} will remain positive wherever {\bb\rr\fb} is positive if
{\bb \calK/\calE^2 \leq 27\mm^2\fb}. In such a case, the null geodesic 
will either be permanently ``incoming'', proceding all the way from
``infinity'' (i.e. the limit {\bb\rr\rightarrow\infty\fb}) down to the
internal singularity (i.e. the limit  {\bb\rr\rightarrow 0\fb}), or else
it will be permanently ``outgoing'', proceding all the way to the
singularity or to infinity depending on whether it inside or outside the
finite horizon radius value {\bb\rr=2\mm\fb} to which it extends in the
infinite past, i.e. as {\bb\uu\rightarrow -\infty\fb}.
As well the such ``ingoing'' and permanently ``outgoing'' possibilities,
the critical case
{\be \calK/\calE^2 = 27\mm^2\label{crit}\fe}
includes also the exceptional possibility of a marginally outgoing --
effectively ``trapped'' -- null trajectory with fixed radius
{\bb \rr=3\mm\, .\fb}

When the angular momentum exceeds this critical value, i.e. if
{\bb \calK/\calE^2 > 27\mm^2\, ,\fb} there will
be a forbidden range {\bb \rr_{-}<\rr<\rr_{+}\fb}
of values of {\bb \rr \fb} for which  {\bb\calR<0\fb}.
It can be seen from \ref{radialf} that relevant limits are explicitly 
obtainable, as the non negative solutions of the cubic equation
{\be \calE^2\rr_{\pm}^3-\calK(\rr_{\pm}-2\mm)=0\, ,\fe}
in the form
{\be \rr_{\pm}=2\sqrt{{\calK}/{3 \calE^2}}\, {\rm cos}\,
\{\psi_{\pm}/3\}\, ,\fe} with
{\be  \psi_{\pm} ={\pi} \mp \,{\rm arcsin}
\sqrt{1-{27\calE^2\mm^2}/{\calK}}\, ,\fe}
which evidently entails the conditions
{\bb 2\mm<\rr_{-}<3\mm<\rr_{+}\, .\fb}

 This means that for a value of  {\bb\calK/\calE^2\fb} above the 
critical bound  (\ref{crit})  the possible null trajectories will be 
classifable as ``free'' or ``trapped''. The ``free'' geodesics are 
initially ``incoming'' from ``infinity'' but  become ``outgoing'' after 
reaching the inner bound at {\bb\rr=\rr_{+}\fb} so as to remain in the
range {\bb\rr\geq\rr_{+}\fb}.  The ``trapped'' geodesics are either
permanently ``ingoing'' within the horizon or else are initially
``outgoing'' from just outside the horizion but become ``ingoing''
after reaching the outer bound  {\bb\rr=\rr_{-}\fb} so as to remain
within the range  {\bb\rr\leq\rr_{-}\fb}.

For a position in the range {\bb \rr \leq 3\mm\fb}  the only kinds of
``bright'' geodesic, meaning those coming from the distant sky at
``infinity'', are of the permanently ``incoming'' kind characterised by
{\bb \calK/\calE^2 \leq 27\mm^2\, .\fb} whereas for
 a position in the range {\bb \rr \leq 3\mm\fb} there will also be
``bright'' geodesics  of the ``free'' kind characterised by
{\bb \calK/\calE^2 > 27\mm^2\, .\fb}
Apart from  the special case of the circular null geodesics at
{\bb \rr=3\mm\, ,\fb} all the other kinds of null geodesic can be
classified as ``dark'' since they can be seen to have
emerged from near the horizon limit radius {\bb \rr\rightarrow 2\mm\fb}
in the distant past (the limit {\bb \uu\rightarrow -\infty)\fb}
and so can be interpreted as trajectories of very highly redshifted
radiation from the infalling matter that be presumed to
originally formed the black hole whose static final state is
under consideration here.

It can be seen that the ratio {\bb\calK/\calE^2\fb} specified as a
function of {\bb {\rm cos}\,\alph\fb} by (\ref{angrat}) will be
monotonically increasing  in the ``incoming'' range, i.e. for
{\bb -1\leq {\rm cos}\,\alph<\sqrt{2\mm/\rr}\fb} where
{\bb \rr>2\mm\fb} and for {\bb -1\leq {\rm cos}\,\alph
<\sqrt{\rr/2\mm}\fb} where {\bb \rr<2\mm\fb}. At the upper end of this
``incoming'' range the ratio {\bb\calK/\calE^2\fb} the tends to a
maximum that will be finite -- with value {\bb \rr^3/(\rr-2\mm)\fb}
-- outside the horizon, but that will be infinite inside the black
hole. The ratio {\bb\calK/\calE^2\fb} will then decrease
monotonically for the higher ``outgoing'' part of the range of
{\bb {\rm cos}\, \alph\, .\fb}

\begin{figure}
\centering
\epsfig{figure=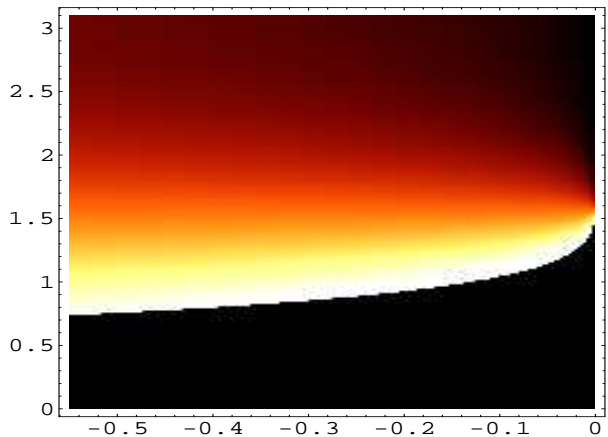, height=6cm, width=8 cm}
\caption{\color{navy}
As in Figure \ref{tfig} but with higher time resolution, showing just
final stages after observer passes  horizon at
{\bb\rr=2\mm\fb} (which occurs when {\bb\tau=-(2/3)^{3/2}\simeq 0.54\fb}
in units used here) as the black hole's apparent angular radius
{\bb\bet\fb} increases towards its upper limit  {\bb\pi/2\fb}
when {\bb\rr\rightarrow 0\fb}.
}
\label{tfin}
\end{figure}

The critical value (\ref{crit}) will be attained for two values of
{\bb {\rm cos}\,\alph\, ,\fb} of which the lower one,
{\bb {\rm cos}\,\alph=\coscr_{_{-}} \fb} say, will be in the
``incoming'' range, and the higher one,
{\bb {\rm cos}\,\alph=\coscr_{_{+}} \fb} will be in the ``outgoing''
range. It can  be seen from (\ref{angrat}) that these values will be
obtainable as the upper and lower roots of 
{\be\rr^2(1-\coscr_{_\pm}^{\,2})=27\mm^2(1-\sqrt{2\mm/\rr}\,
\coscr_{_\pm})^2\, ,\label{cond}\fe}
which will be real and distinct except at {\bb \rr=3\mm\fb}
where they will coincide.The range of angles
characterising the ``bright'' geodesics  will therefore be given by
{\be -1 \leq {\rm cos}\,\alph<{\rm cos}\,\bet\fe} (so that
{\bb \bet \fb} will be interpretable as the apparent angular
radius of the black hole) with a bounding value
{\bb {\rm cos}\,\bet\fb} that will be given by
{\bb {\rm cos}\,\bet=\coscr_{_{-}}\, , \fb} within the radius of
the circular null trajectory, i.e. for {\bb \rr < 3\mm\, ,\fb}
while in the outer regions for which  {\bb \rr > 3\mm\fb}
it will be given by
{\bb {\rm cos}\,\bet=\coscr_{_{+}}\, . \fb}

The required solutions of (\ref{cond}) are 
expressible in terms of the dimensionless variable
{\bb\bar\rr=\rr /2\mm\fb} by
{\bb \coscr_{_\pm}=\left(27\sqrt{\bar\rr} \pm |2\bar\rr^2-3\bar\rr|
\sqrt{\bar\rr^2+3\bar\rr}\right)/(4\bar\rr^3+27) .\fb}
It can thus be seen that (for the freely falling observer) the
apparent angular size {\bb \bet\fb} of the black hole -- as shown in 
the simulation of Figure \ref{whole}, and as plotted against the proper 
time (\ref{propertau}) in  Figure \ref{tfig} and Figure \ref{tfin} --
will be given as a function of the dimensionless radial variable
{\bb \bar\rr=\rr /2\mm\fb} by the analytic formula
{\be{\rm cos}\,\bet=\frac{27\sqrt{\bar\rr}+(2\bar\rr^2-3\bar\rr)
\sqrt{\bar\rr^2+3\bar\rr}}{4\bar\rr^3+27}\, .
\label{wholesize}\fe}.

\end{document}